# Essential role of destructive interference in the gravitationally induced entanglement

Aiham M. Rostom

*Institute of Automation and Electrometry SB RAS, 630090, Novosibirsk, Russia.*[*]

The gravitationally induced entanglement is a type of quantum entanglement that can be generated between two mesoscopic particles using their Newtonian gravitational interaction. It has attracted a great deal of attention as a new platform for studying quantum aspects of gravity. The present paper analyzes the gravitationally induced entanglement as a pure interference effect and shows that the entanglement is induced solely by a sign change associated with the destructive quantum interference. It is also shown that when the entanglement is non-maximal, the preparation for destructive interference for one of the particles can recover a maximum visibility interference pattern for the other particle. Therefore, the non-maximally entangled state can be extremely effective for experimental testing since it can help in reducing requirements (on masses of the particles and their interaction duration, separation distances and sources) and preserve the information about entanglement at the same time. As a result, the improvement in the signal-to-noise ratio is demonstrated and a parameter that determines minimal requirements for experimental testing is defined.

## I. INTRODUCTION

Bose et al.,[1] Marletto and Vedral[2] proposed a method to create entanglement between two mesoscopic particles, each in a superposition of two different locations, by means of their Newtonian gravitational interaction. The claim is that the gravitational field should be a quantum mediator if the entanglement can be verified, what caused a controversy among experts in the quantum gravity.[3–11]

The Bose-Marletto-Vedral (BMV) proposal is based on a previously suggested thought experiment by Richard Feynman, in which he discussed that the explanation of quantum interference between gravitational complex amplitudes must be connected with the quantization of the gravitational field.[12] In BMV proposal, the generation of the entanglement has been explained as being directly related to the complex amplitudes of gravitational interaction. For example, in the ref. [2] one can read: *"...the entanglement between the masses is a function of the relative phase acquired by each of the masses along the paths, via their interaction with the gravitational potential generated by the other superposed mass"*, and in the ref. [1]: *"the quantum mechanical phase induced by their gravitational interaction is significant enough to generate an observable entanglement between the masses"*. In later works, the complex amplitudes are termed as "entanglement phases".[13, 14]

In this work, it is shown that the entanglement is connected not with the gravitational interaction amplitudes but with the type of interference between them, which is a completely different process. In particular, a connection will be drawn between the generation of entanglement and the phase change around the interference minimum.

To this end, the determination of Pancharatnam relative phase (PRP) is an indispensable ingredient. PRP is defined as a total phase (which generally consists of geometrical and dynamical phases) difference between two interfered quantum states.[15, 16] It allows us to find the phase point for which the interference minimum occurs. For instance, let $|a\rangle$ and $|b\rangle$ be two non-orthogonal quantum states, by shifting one of them using a controlled phase shifter $\vartheta$, the interference of their superposition gives

$$\begin{aligned}P &\propto \left(e^{-i\vartheta}\langle a| + \langle b|\right).\left(e^{i\vartheta}|a\rangle + |b\rangle\right) \\ &= \langle a|a\rangle + \langle b|b\rangle + 2|\langle a|b\rangle|\cos\left(\vartheta - \arg\langle a|b\rangle\right),\end{aligned} \quad (1)$$

with $\arg\langle a|b\rangle$ represents the PRP shift of the cosinusoidal variation of $P$ as $\vartheta$ varies. It is clear that the interference is maximum (constructive) for $\vartheta = \arg\langle a|b\rangle$ and minimum (*destructive*) for $\vartheta = \pi + \arg\langle a|b\rangle$. Unlike previous studies which completely ignored the essential role of *destructive* quantum interference in BMV proposal, the present study will lead us also to the optimal experimental arrangements for entanglement observation.

In the next section II, it is shown that the gravitational interactions in BMV experiment shift the interference patterns by PRPs. Using controlled phase shifters, it is possible to compensate the PRPs and arrange for constructive and *destructive* interference. This arrangement leads to *pure* maximally or non-maximally entangled states (according to the strength of the gravitational interaction).

In the section III, we show how the *pure* non-maximally entangled state is effective just as the maximally entangled state in demonstrating the entanglement in BMV proposal. The application of the postselection procedure allows studying the effects of constructive and *destructive* interference separately. Both the probability and phase treatments confirm that the entanglement is induced by the *destructive* quantum interference. Finally, the improvement in the signal-to-noise ratio is demonstrated and a parameter that determines minimal requirements for experimental testing is defined.

Above all, I would like to refer to the recent paper [17] which can provide a photonic quantum simulation for the BMV proposal. It discusses the generation of a pure entangled state between two photons (initially prepared in a product state) using their cross-Kerr interaction in a non-linear optical medium.

## II. QUANTUM INTERFERENCE IN BMV EXPERIMENT

The experimental setup is illustrated in Fig.1. Two test particles of masses $m_1$ and $m_2$ are preselected in the input spatial states $|\psi\rangle = |L\rangle_1 \otimes |R\rangle_2 \equiv |L\rangle_1|R\rangle_2$ of two interferometers. The

---

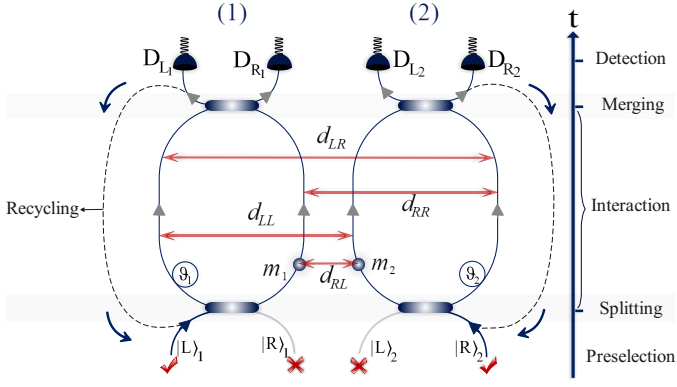

Figure 1. The experimental setup for testing the gravitationally induced entanglement. Two independent mesoscopic particles are preselected in particular spatial states of two interferometers. After the intermediate Newtonian gravitational interactions between particles, a quantum entanglement is induced between the output states of the particles. Essential components in this work are the phase shifters $\vartheta_1$ and $\vartheta_2$ which can be used to control and purify the entanglement. The dashed lines connected the outputs and inputs indicate to the possibility of recycling all unmeasured particles under the postselection on *pure* non-maximally entangled state, see the Discussion section.

splitting and merging of the particles' wave functions are given by the following transformations

$$|R\rangle_\sigma \to \frac{1}{\sqrt{2}}\left(|R\rangle_\sigma + |L\rangle_\sigma\right), \qquad \sigma = 1, 2 \quad (2)$$

$$|L\rangle_\sigma \to \frac{1}{\sqrt{2}}\left(|R\rangle_\sigma + e^{i\pi}|L\rangle_\sigma\right), \quad (3)$$

and the state of the composite system after passing the input beam splitters becomes

$$|\psi\rangle \to \frac{1}{2}(|R\rangle_1 - |L\rangle_1) \otimes (|R\rangle_2 + |L\rangle_2)$$
$$= \frac{1}{2}(|R\rangle_1|R\rangle_2 + |R\rangle_1|L\rangle_2 - |L\rangle_1|R\rangle_2 - |L\rangle_1|L\rangle_2). \quad (4)$$

In the Newtonian regime, the Schrödinger equation takes the simple form

$$i\hbar\frac{d|\psi\rangle}{dt} \approx \widehat{H}_{ij}|\psi\rangle, \quad \widehat{H}_{ij} = -\frac{Gm_1m_2}{\widehat{d}_{ij}}, \quad (5)$$

here $\widehat{H}_{ij}$ is the interaction Hamiltonian,[18, 19] $G$ is the gravitational constant, $d_{ij}$ is the distance between the particles as shown in Fig.1. Thus, under the gravitational interaction, the branches of the wave functions are transformed according to the following time-evolution operator

$$\widehat{U}_{ij} = \exp\left(-i\frac{\widehat{H}_{ij}\tau}{\hbar}\right) = \exp\left(i\frac{Gm_1m_2\tau}{\hbar \widehat{d}_{ij}}\right) = \exp(i\widehat{\phi}_{ij}), \quad (6)$$

where $\tau$ is the interaction duration. To control the interference and measure the shift of the output interference pattern, controlled phase shifts $\vartheta_1$ and $\vartheta_2$ are applied on one branch of the wave functions of each particle $\widehat{U}_1|L\rangle_1 \to$ $\exp(i\vartheta_1)|L\rangle_1$, $\widehat{U}_2|R\rangle_2 \to \exp(i\vartheta_2)|R\rangle_2$.[20] The operators $\widehat{U}_{ij}$, $\widehat{U}_1$ and $\widehat{U}_2$ act on the state $|\psi\rangle$ and just before entering the final beam splitters transform it into

$$|\psi\rangle \to \frac{1}{2}\left(\widehat{U}_2\widehat{U}_{RR}|R\rangle_1|R\rangle_2 + \widehat{U}_{RL}|R\rangle_1|L\rangle_2 \right.$$
$$\left. -\widehat{U}_1\widehat{U}_2\widehat{U}_{LR}|L\rangle_1|R\rangle_2 - \widehat{U}_1\widehat{U}_{LL}|L\rangle_1|L\rangle_2\right)$$
$$= \frac{1}{2}\left(e^{i\vartheta_2}e^{i\phi_{RR}}|R\rangle_1|R\rangle_2 + e^{i\phi_{RL}}|R\rangle_1|L\rangle_2 \right.$$
$$\left. -e^{i\vartheta_1}e^{i\vartheta_2}e^{i\phi_{LR}}|L\rangle_1|R\rangle_2 - e^{i\vartheta_1}e^{i\phi_{LL}}|L\rangle_1|L\rangle_2\right). \quad (7)$$

Finally, the superposition components merge in the output beam splitters according to the transformations (2) and (3)

$$|\psi\rangle \to \frac{1}{4}\left[(|R\rangle_1 + |L\rangle_1)(|R\rangle_2 + |L\rangle_2)e^{i\vartheta_2}e^{i\phi_{RR}} \right.$$
$$+ (|R\rangle_1 + |L\rangle_1)(|R\rangle_2 - |L\rangle_2)e^{i\phi_{RL}}$$
$$- (|R\rangle_1 - |L\rangle_1)(|R\rangle_2 + |L\rangle_2)e^{i\vartheta_1}e^{i\vartheta_2}e^{i\phi_{LR}}$$
$$\left. - (|R\rangle_1 - |L\rangle_1)(|R\rangle_2 - |L\rangle_2)e^{i\vartheta_1}e^{i\phi_{LL}}\right], \quad (8)$$

and the two-particle output state is given by

$$|\Psi\rangle = \alpha|R\rangle_1|R\rangle_2 + \beta|R\rangle_1|L\rangle_2 + \gamma|L\rangle_1|R\rangle_2 + \delta|L\rangle_1|L\rangle_2, \quad (9)$$

which is a general entangled state with the following complex amplitudes

$$\alpha = \left[e^{i(\vartheta_2+\phi_{RR})} + e^{i\phi_{RL}} - e^{i(\vartheta_1+\vartheta_2+\phi_{LR})} - e^{i(\vartheta_1+\phi_{LL})}\right]/4,$$
$$\beta = \left[e^{i(\vartheta_2+\phi_{RR})} - e^{i\phi_{RL}} - e^{i(\vartheta_1+\vartheta_2+\phi_{LR})} + e^{i(\vartheta_1+\phi_{LL})}\right]/4,$$
$$\gamma = \left[e^{i(\vartheta_2+\phi_{RR})} + e^{i\phi_{RL}} + e^{i(\vartheta_1+\vartheta_2+\phi_{LR})} + e^{i(\vartheta_1+\phi_{LL})}\right]/4,$$
$$\delta = \left[e^{i(\vartheta_2+\phi_{RR})} - e^{i\phi_{RL}} + e^{i(\vartheta_1+\vartheta_2+\phi_{LR})} - e^{i(\vartheta_1+\phi_{LL})}\right]/4.$$

The output probabilities of detecting the particle 1 in a particular spatial state is then given by

$$P_{R_1} = {}_1\langle R|\left[\text{Tr}_2(|\Psi\rangle\langle\Psi|)\right]|R\rangle_1 = \alpha\alpha^* + \beta\beta^*$$
$$= \frac{1}{2}\left[1 + v\cos(\pi + \vartheta_1 - \Delta_1)\right], \quad (10)$$
$$P_{L_1} = {}_1\langle L|\left[\text{Tr}_2(|\Psi\rangle\langle\Psi|)\right]|L\rangle_1 = \gamma\gamma^* + \delta\delta^*$$
$$= \frac{1}{2}\left[1 + v\cos(\vartheta_1 - \Delta_1)\right], \quad (11)$$

where $v = \cos(\xi/2)$ is the visibility of the interference pattern with $\xi = \phi_{RR} - \phi_{RL} - \phi_{LR} + \phi_{LL}$, and $\Delta_1 = (\phi_{RR} + \phi_{RL} - \phi_{LR} - \phi_{LL})/2$, is the PRP shift of the interference pattern. In interferometer 2, we have the detection probabilities

$$P_{R_2} = {}_2\langle R|\left[\text{Tr}_1(|\Psi\rangle\langle\Psi|)\right]|R\rangle_2 = \alpha\alpha^* + \gamma\gamma^*$$
$$= \frac{1}{2}\left[1 + v\cos(\vartheta_2 - \Delta_2)\right], \quad (12)$$
$$P_{L_2} = {}_2\langle L|\left[\text{Tr}_1(|\Psi\rangle\langle\Psi|)\right]|L\rangle_2 = \beta\beta^* + \delta\delta^*$$
$$= \frac{1}{2}\left[1 + v\cos(\pi + \vartheta_2 - \Delta_2)\right], \quad (13)$$

with the same visibility $v$ but a different PRP shift of the interference pattern $\Delta_2 = (-\phi_{RR} + \phi_{RL} - \phi_{LR} + \phi_{LL})/2$.

Formulae (10), (11), (12) and (13) describe a statistical ensemble of infinitely repeated observations. In each interferometer, as it has been mentioned in the introduction, we can distinguish between two main types of interference, constructive and *destructive*. For instance, in the interferometer 2, by setting the controlled phase shifter $\vartheta_2 = \Delta_2$, one of the outputs will be characterized by a constructive interference with the highest possible probability $P_{R_2}$, while the second output by a *destructive* interference with the lowest possible probability $P_{L_2}$.

To turn the entanglement (9) into its pure form (EPR state), one can prepare for the aforementioned types of interference by carefully adjusting the controlled phase shifters to the values $\vartheta_1 = \Delta_1, \vartheta_2 = \Delta_2$. The probability amplitudes $(\alpha, \beta, \gamma, \delta)$ then become

$$\alpha = 0, \ \beta = \left[e^{\frac{i}{2}(\phi_{RR}+\phi_{RL}-\phi_{LR}+\phi_{LL})} - e^{i\phi_{RL}}\right]/2,$$
$$\delta = 0, \ \gamma = \left[e^{i\phi_{RL}} + e^{\frac{i}{2}(\phi_{RR}+\phi_{RL}-\phi_{LR}+\phi_{LL})}\right]/2.$$

Consequently, links $|R\rangle_1|R\rangle_2$ and $|L\rangle_1|L\rangle_2$ vanish and the output entangled state (9) turns into

$$|\Psi\rangle = i\sin\left(\frac{\xi}{4}\right)|R\rangle_1|L\rangle_2 + \cos\left(\frac{\xi}{4}\right)|L\rangle_1|R\rangle_2, \quad (14)$$

which is a *pure* non-maximally entangled states for $\xi \neq \pi$, up to an unimportant overall phase factor. If the interactions are strong enough to produce the value $\xi = \pi$, the entangled state becomes

$$|\Psi\rangle = \frac{1}{\sqrt{2}}\left(i|R\rangle_1|L\rangle_2 + |L\rangle_1|R\rangle_2\right), \quad (15)$$

which is a *pure* maximally entangled state. Note that the output states characterized by the same type of interference are linked together.

## III. DISCUSSION

From now on, we assume that $\phi_{RL} \gg (\phi_{LL}, \phi_{RR}, \phi_{LR})$ and $\xi \simeq -\phi_{RL} \equiv -\phi$ is weak. This gives the following *pure* non-maximally entangled state

$$|\Psi\rangle = -i\sin\left(\frac{\phi}{4}\right)|R\rangle_1|L\rangle_2 + \cos\left(\frac{\phi}{4}\right)|L\rangle_1|R\rangle_2. \quad (16)$$

The most important feature of this state is that it contains a maximal correlation between rare events (in $R_1$ and $L_2$ outputs due to the *destructive* interference). *Rare* events are intuitively more informative events than high-probability ones. Shannon defined this fact by introducing a simple quantity called information content: $I = -\log P$, which measures the unexpectedness of an event that occurs with probability $P$.[21]

Apart from the high information content, the *pure* non-maximally entangled state helps in reducing the requirements on the masses of the particles and their interaction time, separation distances and sources, and consequently simplifies the experimental observation of the entanglement.

### A. Recovering a maximum visibility interference pattern

Now let us suppose that interferometer 2 is slightly smaller in size than interferometer 1, so that the quantum interference in the former precedes in time the interference in the latter.

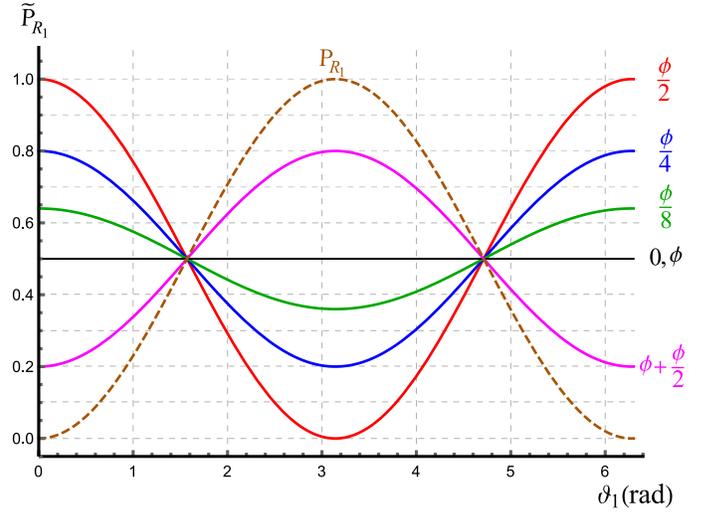

Figure 2. The probability $\widetilde{P}_{R_1}$ versus the controlled phase shift $\vartheta_1$ for $\phi = 10^{-4} rad$ and different values of $\vartheta_2$. The dashed curve represents the interference pattern described by $P_{R_1}$. Inside the interval $0 < \vartheta_2 < \phi$, the interference pattern given by $\widetilde{P}_{R_1}$ is shifted from $P_{R_1}$ by a $\pi$ phase, the visibility $\widetilde{v}$ becomes maximum for $\vartheta_2 = \phi/2$ and vanishes completely for $\vartheta_2 = 0$ or $\phi$. Outside this interval, $P_{R_1}$ and $\widetilde{P}_{R_1}$ oscillate in phase with each other.

In the output of interferometer 2, we apply the postselection procedure and study the corresponding behavior in interferometer 1. Let the postselected state be $|L\rangle_2$, then the state $|R\rangle_2$ must be discarded from the output wave function (9)

$$|\Psi\rangle = \alpha|R\rangle_1\cancel{|R\rangle_2} + \beta|R\rangle_1|L\rangle_2 + \gamma|L\rangle_1\cancel{|R\rangle_2} + \delta|L\rangle_1|L\rangle_2$$
$$\to \beta|R\rangle_1|L\rangle_2 + \delta|L\rangle_1|L\rangle_2 \propto |\widetilde{\Psi}\rangle, \quad (17)$$

and the resultant postselected state $|\widetilde{\Psi}\rangle$ is not normalized. By normalization

$$|\widetilde{\Psi}\rangle = \frac{1}{\sqrt{\beta\beta^* + \delta\delta^*}}(\beta|R\rangle_1 + \delta|L\rangle_1)|L\rangle_2 = |\widetilde{\Phi}\rangle_1|L\rangle_2, \quad (18)$$

where $\langle\widetilde{\Psi}|\widetilde{\Psi}\rangle = \sqrt{\beta\beta^* + \delta\delta^*}$. With every successful postselection in interferometer 2, the corresponding detection probabilities in interferometer 1 are given by

$$\widetilde{P}_{R_1} = \frac{\beta\beta^*}{\beta\beta^* + \delta\delta^*} = \frac{1}{2}\left[1 + \widetilde{v}\cos\left(\pi + \vartheta_1 - \frac{\phi}{2}\right)\right], \quad (19)$$
$$\widetilde{P}_{L_1} = 1 - \widetilde{P}_{R_1}, \quad (20)$$

where

$$\widetilde{v} = \frac{2\sin\frac{\vartheta_2}{2}\sin\left(\frac{\vartheta_2-\phi}{2}\right)}{1 - \cos\left(\frac{\phi}{2}\right)\cos\left(\vartheta_2 - \frac{\phi}{2}\right)}, \quad (21)$$

is the visibility of interference pattern of particle 1 under the postselection on the quantum states of particle 2. It depends not only on $\phi$ but also on the dynamical phase $\vartheta_2$.

At first sight, one may think that there is no difference between cosinusoidal variations of $P_{R_1}$ and $\widetilde{P}_{R_1}$ as a function of $\vartheta_1$, both vary in phase with each other and both describe a *destructive* interference for $\vartheta_1 = \phi/2$. This is true except in the interval $0 < \vartheta_2 < \phi$ in which the visibility $\widetilde{\nu}$ takes negative values and the interference described by $\widetilde{P}_{R_1}$ is shifted by a $\pi$ phase and becomes constructive.[22] The visibility $\widetilde{\nu}$ in this interval can range from 0 (no interference) when $\vartheta_2 = 0$ or $\phi$, to $-1$ (maximum interference) when $\vartheta_2 = \phi/2$. Negative visibility means a phase reversal as it is shown in Fig.2.

Unlike $P_{R_1}$, the probability $\widetilde{P}_{R_1}$ depends on $\vartheta_2$. This $\vartheta_2$-dependence is a hidden interference phenomenon that can be recovered by the postselection,[23, 24] with a visibility of $\cos(\vartheta_1 - \phi/2)$. Fig.3, in which the probability $\widetilde{P}_{R_1}$ is drawn as a function of $\vartheta_2$, shows that this recovered interference has maximum visibility for weak interaction strengths. The output spatial modes of interferometer 1 are completely switched with every *destructive* interference in interferometer 2.

In the literature, most attention has been concentrated on increasing the degree of entanglement, assuming that if the gravitationally induced phase is weak, the quantum interference of masses is negligible and not observable.[1, 2, 8, 13] In contrast, the present results shows that for small gravitational phases $\phi \ll \pi$, what we only need to recover a maximum visibility interference pattern is a good control over the phase $\vartheta_2$.

### B. Sign change due to the destructive interference

When the interference described by the output state $|L\rangle_2$ takes place (successful postselection), the internal state of particle 1 before it enters the output beam splitter becomes

$$
\begin{aligned}
|\widetilde{r}\rangle_1 &\propto {}^{post}(_2\langle L|)\widehat{U}_1\widehat{U}_2\widehat{U}_{RL}\left(|L\rangle_1|R\rangle_2\right)^{pre}\\
&\to {}^{post}(_2\langle L|)\big[|R\rangle_1(|R\rangle_2+|L\rangle_2)e^{i\vartheta_2}\\
&\quad +|R\rangle_1(|R\rangle_2-|L\rangle_2)e^{i\phi}\\
&\quad -|L\rangle_1(|R\rangle_2+|L\rangle_2)e^{i\vartheta_1}e^{i\vartheta_2}\\
&\quad -|L\rangle_1(|R\rangle_2-|L\rangle_2)e^{i\vartheta_1}\big]\\
&= \left(e^{i\vartheta_2}-e^{i\phi}\right)|R\rangle_1 + e^{i\pi}\left(e^{i\vartheta_2}-1\right)e^{i\vartheta_1}|L\rangle_1, \quad (22)
\end{aligned}
$$

where the superscripts (pre) and (post) stand for the preselected and postselected states, respectively. We set $\vartheta_1 = \phi/2$. For $\vartheta_2 = \pi + \phi/2$, the postselection on $|L\rangle_2$ results from constructive interference and

$$|\widetilde{r}\rangle_1 \propto \frac{e^{i\frac{\phi}{2}}+e^{i\phi}}{\sqrt{2}}\left(|R\rangle_1 + e^{i\pi}|L\rangle_1\right),$$

the internal state of particle 1 is superposed with $(-)$ phase as expected from (3). For $\vartheta_2 = \phi/2$, the postselection on $|L\rangle_2$ results from *destructive* interference and

$$|\widetilde{r}\rangle_1 \propto \frac{e^{i\frac{\phi}{2}}-e^{i\phi}}{\sqrt{2}}\left(|R\rangle_1 + |L\rangle_1\right),$$

the internal state of particle 1 is superposed with $(+)$ phase. Accordingly, regardless of the gravitational interaction strength, the information flows between the entangled particles as a sign change in the internal quantum state of interferometer 1 due to the occurrence of *destructive* interference in interferometer 2.

This gives rise to the following question: is there any relation between the sign change and the quantization of the gravitational field?. The answer is likely "no". The gravitational interaction may be extremely weak (or extremely strong, slightly less than $2\pi$) in which its strength is not enough (or suitable) to produce a sign change.

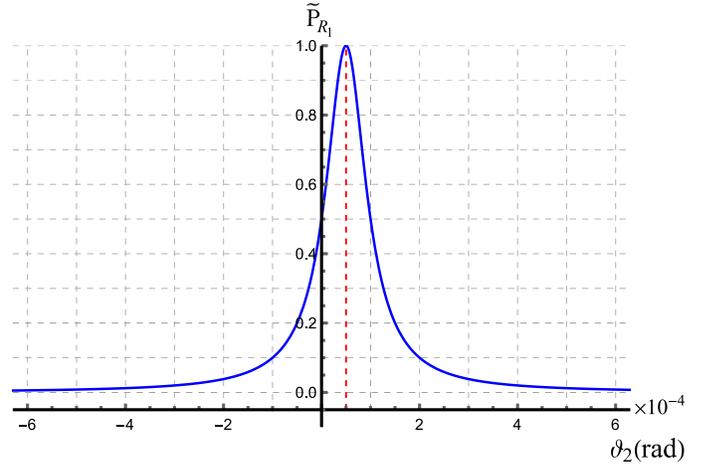

Figure 3. The probability $\widetilde{P}_{R_1}$ as a function of $\vartheta_2$ for $\phi = 10^{-4}rad$ and $\vartheta_1 = \phi/2$. With every *destructive* interference of wave function of the particle 2, a phase-like variable of $\pi$ switches the outputs of the interferometer 1. This phase can be estimated experimentally from the formula: $\arccos\left[\left(\widehat{P}_{L_1}-\widehat{P}_{R_1}\right)/\left(\widehat{P}_{L_1}+\widehat{P}_{R_1}\right)\right]$.[25] The visibility of this interference is $\cos(\vartheta_1 - \phi/2) = 1$. By keeping $\vartheta_1 \ll \pi/2$, the visibility will be close to unity regardless of how small the interaction strength $\phi$ is, and for large values of $\phi$ it decreases until the interference disappears completely for $\phi = \pi$.

Moreover, we need to keep in mind that the *destructive* interference leads to phase changes which are highly non-intuitive and hard to interpret. This includes discontinuous phase jumps or rapid and smooth changes in phases.[26–30] The entanglement in BMV experiment is not unrelated to phase changes associated with the destructive interference.[31] Further investigation into the geometrical and topological descriptions of such phase changes is needed for a more comprehensive understanding.

### C. Parameter for minimal experimental requirements

The measurement signal-to-noise ratio (SNR) is defined as the ratio of estimated quantity to its standard deviation. The quantity of interest here is the total number of particles $N_{R_1}$ that leave interferometer 1 from $R_1$ port for $N_{L_2}$ successful postselections in interferometer 2. Obviously, $N_{R_1} \le N_{L_2}$ is a random variable that obeys the binomial distribution, and thus we can model the expected SNR as

$$SNR = \frac{\langle N_{R_1}\rangle}{\sigma(N_{R_1})} = \frac{N_{L_2}\widetilde{P}_{R_1}}{\sqrt{N_{L_2}\widetilde{P}_{R_1}(1-\widetilde{P}_{R_1})}} = \sqrt{N_{L_2}\frac{\widetilde{P}_{R_1}}{\widetilde{P}_{L_1}}}, \quad (23)$$

with $\langle N_{R_1}\rangle$ the average number of particles detected in the output $R_1$, $\sigma$ is its standard deviation.

For careful adjustment of $\vartheta_1$ and $\vartheta_2$ near the optimal point of operation $\phi/2$, the state of the particles is described by the *pure* non-maximally entangled state (16). However, with $N_{L_2}$ postselected particles, it is unsurprising that experimentalists will never be able to detect all partner particles in the output $R_1$ or justify asserting that the detection in the another output $L_1$ will never occur for a larger sample size.

Therefore, under the best experimental conditions, a large number of postselections $N_{L_2}$ can yield frequencies with at least $N_{L_1} = 1$, $N_{R_1} = N_{L_2} - 1$. From these frequencies, the probabilities $\widehat{P}_{R_1} = N_{R_1}/N_{L_2}$ and $\widehat{P}_{L_1} = N_{L_1}/N_{L_2}$ are inferred. The observed SNR is

$$SNR = \sqrt{N_{L_2} \frac{\widehat{P}_{R_1}}{\widehat{P}_{L_1}}} = \sqrt{N_{L_2} \frac{N_{R_1}}{N_{L_1}}} \lesssim N_{L_2}. \qquad (24)$$

This is the highest SNR that we can get. In the opposite case when $N_{L_1} = N_{L_2} - 1$, $N_{R_1} = 1$, the $SNR \gtrsim 1$. Thus, the postselection on the non-maximally entangled state can help in improving the SNR, keeping in mind that the detection events in $L_2$ and $R_1$ are rare by assumption. In other words, almost no simultaneous detection event in ports $L_2$ and $R_1$ can result from a coincidence with any noise factor but it is a signal comes from the correlation between these rare events.

Now if $T$ is the duration of the experiment and $\Gamma$ is the input rate of the particles (pairs), the number of postselections becomes $N_{L_2} = P_{L_2} \Gamma T$. This leads us to define the condition for testing BMV proposal as

$$N_{L_2} \simeq \left(\frac{\phi}{4}\right)^2 \Gamma T \gg 1 \implies \Gamma T \left(\frac{G m_1 m_2 \tau}{4 d \hbar}\right)^2 \gg 1,$$

where $P_{L_2} = \sin^2 \frac{\phi}{4} \simeq \left(\frac{\phi}{4}\right)^2$, $d \equiv d_{RL}$, and the "much greater" symbol is used since a statistical ensemble is needed in realistic experiments, i.e. $N_{L_2} \gg 1$ (say by a factor of $10^2$ or more). Therefore, any experimental parameter $\kappa$ that satisfies the following inequality

$$\kappa = \Gamma T \left(\frac{m_1 m_2 \tau}{d}\right)^2 \gg \frac{16 \hbar^2}{G^2} = 4 \times 10^{-47} \text{ kg}^4 \cdot \text{m}^{-2} \cdot \text{s}^2, \qquad (25)$$

will allow the testing of BMV proposal no matter how small the gravitational phase is.

Note that the combination between the postselection and the *pure* non-maximally entangled state provides an additional technical advantage. The postselection does not mean to throw away the large amount of unpostselected particles. All of them can be reinjected into the interferometer until they get successfully postselected, as is illustrated in the Fig.1. This helps in increasing the input rate of the particles $\Gamma$ drastically, taking into account that the time delay between successive injection of a pair of particles must be greater than the temporal width of the pair's wave packet.

In addition to what we have already seen here, the *pure* non-maximally entangled state has been shown to be more efficient than the maximally entangled state in many applications. It can lead to the maximal violation of several Bell inequalities[32] or help in: reducing of the required detector efficiency for a loophole-free test of quantum nonlocality,[33, 34] the simulation of entanglement,[35] increasing the efficiency in quantum cryptography,[36] optimal quantum estimation of phases,[17] and the complete and perfect quantum teleportation.[37] Not to forget that the non-maximally entangled state is also easier to prepare in the lab than the maximally entangled one.

Finally, I would like to end with two quotations from two great physicists Feynman and Dirac to emphasize more on the necessity for further investigation into the nature of interference, phases and phase factors (to reveal the underlying mechanism of quantum entanglement). Richard Feynman in his famous lectures described the interference in the double-slit experiment as[38]

> ...a phenomenon which is impossible, absolutely impossible, to explain in any classical way, and which has in it the heart of quantum mechanics. In reality, it contains the only mystery.

Paul Dirac pointed out the intrinsic role of complex phase factors in the interference phenomena, he said:[39]

> So if one asks what is the main feature of quantum mechanics, I feel inclined now to say that it is not non-commutative algebra. It is the existence of probability amplitudes which underlie all atomic processes. Now a probability amplitude is related to experiment but only partially. The square of its modulus is something that we can observe. That is the probability which the experimental people get. But besides that there is a phase, a number of modulus unity which can modify without affecting the square of the modulus. And this phase is all important because it is the source of all interference phenomena, but its physical significance is obscure. So the real genius of Heisenberg and Schrödinger, you might say, was to discover the existence of probability amplitudes containing this phase quantity which is very well hidden in nature, and it is because it was so well hidden that people hadn't thought of quantum mechanics much earlier.

## IV. CONCLUSION

In conclusion, the *destructive* quantum interference plays an essential role in interpretation and experimental observation of the gravitationally induced entanglement.

The *destructive* interference between the superposition components of one particle induces a sign change (a relative $\pi$ phase factor) in the internal quantum state of the second particle. This sign change is the "amplitude" that lies behind the entanglement between the particles.

In the weak coupling regime, the application of the postselection procedure on the quantum states of one of the particles and arrangement for *destructive* interference using the controlled phase shifter can recover a maximum visibility interference pattern for the second particle. From the recovered interference, the entanglement can be observed.

In the end, the entanglement as a "$\pi$ phase factor effect" invites us to take a step back and search for the possible geometrical and topological properties of the quantum space. We can then judge whether the geometry of space originates from the quantum *or vice versa*.

## ACKNOWLEDGMENTS

The author thanks Prof. Leonid Il'ichov for valuable discussions and for reading the draft of this paper and providing very helpful feedback. This work was performed at the Institute of Automation and Electrometry of the Siberian Branch of the Russian Academy of Sciences within the framework of the State Assignment (Project No. AAAA-A21-121021800168-4).

[1] S. Bose, A. Mazumdar, G. W. Morley, H. Ulbricht, M. Toroš, M. Paternostro, A. A. Geraci, P. F. Barker, M. S. Kim, and G. Milburn, Phys. Rev. Lett. **119**, 240401 (2017).

[2] C. Marletto and V. Vedral, Phys. Rev. Lett. **119**, 240402 (2017).

[3] E. Martín-Martínez and T. R. Perche, arXiv:2208.09489 [quant-ph] (2022).

[4] M. Christodoulou, A. Di Biagio, M. Aspelmeyer, Č. Brukner, C. Rovelli, and R. Howl, arXiv:2202.03368 [quant-ph] (2022).

[5] D. Carney, Y. Chen, A. Geraci, H. Müller, C. D. Panda, P. C. Stamp, and J. M. Taylor, arXiv:2203.11846 [gr-qc] (2022).

[6] C. Anastopoulos, M. Lagouvardos, and K. Savvidou, Class. Quantum Grav. **38**, 155012 (2021).

[7] T. D. Galley, F. Giacomini, and J. H. Selby, arXiv:2012.01441 [quant-ph] (2020).

[8] M. Christodoulou and C. Rovelli, Phys. Lett. B **792**, 64 (2019).

[9] M. J. W. Hall and M. Reginatto, J. Phys. A: Math. Theor. **51**, 085303 (2018).

[10] N. Altamirano, P. Corona-Ugalde, R. B. Mann, and M. Zych, Class. Quantum Grav. **35**, 145005 (2018).

[11] C. Anastopoulos and B.-L. Hu, arXiv:1804.11315 [quant-ph] (2018).

[12] R. Feynman, in *Chapel Hill Conference Proceedings* (1957).

[13] T. W. van de Kamp, R. J. Marshman, S. Bose, and A. Mazumdar, Phys. Rev. A **102**, 062807 (2020).

[14] R. J. Marshman, A. Mazumdar, and S. Bose, Phys. Rev. A **101**, 052110 (2020).

[15] A. Shapere and F. Wilczek, *Geometric phases in physics* (World scientific, 1989).

[16] E. Sjöqvist, A. K. Pati, A. Ekert, J. S. Anandan, M. Ericsson, D. K. L. Oi, and V. Vedral, Phys. Rev. Lett. **85**, 2845 (2000).

[17] A. M. Rostom, Ann. Phys. (Berlin) **534**, 2100434 (2022).

[18] H. C. Nguyen and F. Bernards, The European Physical Journal D **74**, 1 (2020).

[19] D. Miki, A. Matsumura, and K. Yamamoto, Phys. Rev. D **103**, 026017 (2021).

[20] Note that these phase shifts can be generated also by placing particles of masses $m$ (each not in a superposed state) near the internal branches $L_1$ and $R_2$ of the interferometers, or simply by varying the branches' lengths.

[21] K. Sayood, *Introduction to Data Compression* (Morgan Kaufmann, Burlington, 2018).

[22] If it is possible to regard the *destructive* interference of particle 2 as a loss of phase information in the output $L_2$, then this information cannot be lost, it is conserved because it is recovered by constructive interference of particle 1 in the output $R_1$.

[23] T. Yakovleva, A. Rostom, V. Tomilin, and L. V. Il'ichov, Quantum Electron. **49**, 439 (2019).

[24] T. S. Yakovleva, A. M. Rostom, V. A. Tomilin, and L. V. Il'ichov, Quantum Stud.: Math. Found. **6**, 217 (2019).

[25] L. Pezzé, A. Smerzi, G. Khoury, J. F. Hodelin, and D. Bouwmeester, Phys. Rev. Lett. **99**, 223602 (2007).

[26] R. Bhandari, Phys. Rep. **281**, 1 (1997).

[27] S. Tamate, H. Kobayashi, T. Nakanishi, K. Sugiyama, and M. Kitano, New J. Phys. **11**, 093025 (2009).

[28] R. M. Camacho, P. B. Dixon, R. T. Glasser, A. N. Jordan, and J. C. Howell, Phys. Rev. Lett. **102**, 013902 (2009).

[29] N. E. Huang, S. R. Long, and Z. Shen (Elsevier, 1996).

[30] G. J. Gbur, *Singular Optics* (CRC Press, 2017).

[31] This is clear from the visibility in Eq. (21) which changes its sign in the interval $0 < \vartheta_2 < \phi$. The phase of visibility jumps from 0 to $\pi$.

[32] N. Brunner, D. Cavalcanti, S. Pironio, V. Scarani, and S. Wehner, Rev. Mod. Phys. **86**, 419 (2014).

[33] P. H. Eberhard, Phys. Rev. A **47**, R747 (1993).

[34] B. G. Christensen, K. T. McCusker, J. B. Altepeter, B. Calkins, T. Gerrits, A. E. Lita, A. Miller, L. K. Shalm, Y. Zhang, S. W. Nam, N. Brunner, C. C. W. Lim, N. Gisin, and P. G. Kwiat, Phys. Rev. Lett. **111**, 130406 (2013).

[35] N. Brunner, N. Gisin, and V. Scarani, New J. Phys. **7**, 88 (2005).

[36] P. Xue, C.-F. Li, and G.-C. Guo, Phys. Rev. A **64**, 032305 (2001).

[37] M. Ohya and I. Volovich, *Mathematical foundations of quantum information and computation and its applications to nano-and bio-systems* (Springer, 2011).

[38] R. P. Feynman, R. B. Leighton, and M. Sands, *The Feynman lectures on physics* (Addison–Wesley, 1963).

[39] C. N. Yang, Int. J. Mod. Phys. A **18**, 3263 (2003).